\documentclass{appolb}
\usepackage{graphicx}
\usepackage{bm}
\usepackage{cite}

\def\be{\begin{equation}}
\def\ee{\end{equation}}
\def\bea{\begin{eqnarray}}
\def\eea{\end{eqnarray}}

\def\ptl{\partial}
\begin{document}
\title{\bf Equation of a vacuum state and a structure formation in the
universe}%

\author{{S.L.Cherkas
\address{Institute for Nuclear Problems, Bobruiskaya
11, Minsk 220030, Belarus \\$~$ }}
\\
{V.L. Kalashnikov
\address{ La Sapienza Universit\'a di Roma, Via
Eudossiana 18, 00189 - Roma, RM, Italia}}}

\maketitle
\begin{abstract}
The vacuum is considered as some fluid emergent from the
zero-point fluctuations of the quantum fields contributing to the
vacuum energy density and pressure. The equation of vacuum state
and the speed of vacuum sound-waves are deduced under the
assumption of zero vacuum entropy. The evolution of the background
space-time metric resembles that of the Milne's-like universe. In
the framework of the five-vector theory of gravitation allowing an
arbitrary choice of the energy density reference level, the
dynamics of the vacuum, pressureless matter, and space-time
metrics perturbations are traced under this background. The
obtained results show the very early formation of the Universe
structure without the need for dark matter. Thus, a vacuum can be
considered as some of the dark-energy-matter unification.
\end{abstract}

\section{Introduction}

Clarification of the role of vacuum in the formation of the cosmic
microwave background anisotropy (CMB) and the matter structures in
the evolving Universe remains one of the key issues of modern
cosmology despite the numerous hypotheses and suggestions
\cite{Zeldovich1981,Weinberg1989,Sahni2000,Carroll2001,Padmanabhan2003,Chernin2008,Li2011}.
At the same time, the application of the standard renormalization
procedure \cite{Birrell1982} to gravitation seems not feasible due
to the impossibility to define a vacuum state which is invariant
under the general coordinate transformations \cite{Birrell1982,
Padmanabhan2003,anischenko2008functional}. Nevertheless, one might
intuitively feel that some ``pieces'' of the vacuum energy density
and pressure have to be omitted, while the others are to be taken
into account as was demonstrated on the example of the Gowdy's
model \cite{Cherkas2017}. The issue of huge vacuum energy
indicates that the most diverging part of the vacuum energy
density has no physical meaning and has to be discarded. Formally,
this is impossible in the frameworks of the general theory of
relativity (GR) because of any non-zero energy density contributes
to space-time curvature. However, this procedure guaranteeing
against the unphysical ``piece'' of the vacuum energy density can
be realized within the so-called five-vector theory of gravitation
(FVT)  \cite{Cherkas2019}, in which the energy density is defined
up to some constant. The remaining part of the vacuum energy can
be treated as corresponding to some``fluid'' possessing definite
equation of state (ES).

The idea to describe a vacuum as some ``fluid'' defined by ES
seems very tempting, starting from the concepts of a quantum
``ether'' \cite{Dirac1951,Zolotarev1985} to the models of
quintessence, K-essence, and cosmological Chaplygin's gas
\cite{Bento2002,Silva2003,Amendola2010}. Generally speaking, the
the situation looks as follows: there is no ``ether'' in the flat
space-time owing to vacuum invariance relatively the Lorentz
transformations, but in the presence of gravitation, the picture
is different because there is no invariant vacuum state relatively
the general coordinate transformations. As a result, one might
conjecture the existence of some preferred reference frame
indicating the existence of ``ether'' identified with a quantum
vacuum.

\section{Violation of gauge invariance in a framework of FVT}

 In GR,
any spatially uniform energy density (including that of zero-point
fluctuations of the quantum fields) causes the expansion of the
universe. Using the Planck level of UV-cutoff results in the
Planckian vacuum energy density $\rho_{vac}\sim M_p^4$
\cite{Weinberg1989}, which leads to the universe expanding with
the Planckian rate \cite{d1}. In this sense, the vacuum energy
problem is an observational fact \cite{arx}. One of the possible
solutions is to build a theory of gravity, allowing an arbitrarily
reference level of energy density. One such theory has long been
known. That is the unimodular gravity \cite{16,17,21,22,23,unim},
which admits an arbitrary cosmological constant. However, under
using of the comoving momentums cutoff, the vacuum energy density
scales with time as radiation \cite{Visser18,arx}, but not as the
cosmological constant.

Recently, another theory has been suggested \cite{Cherkas2019},
which also leads to the Friedman equation defined up to some
arbitrary constant. This constant corresponds to the invisible
radiation and, thus, can compensate the vacuum energy.

Five-vector theory of gravity (FVT) \cite{Cherkas2019} assumes the
gauge invariance violation in GR  by constraining the class of all
possible metrics in varying the standard Einstein-Hilbert action.
This theory arises if one varies the standard Einstein-Hilbert
action over not all possible space-time metrics $g_{\mu\nu}$, but
over some class of conformally-unimodular metrics\footnote{In this
gauge, a space-time metric is presented as a product of a common
multiplier by a 4-dimensional matrix with a determinant equal to
-1, including a 3-dimensional spatial block with unit
determinant.} \cite{Cherkas2019}
\be ds^2\equiv g_{\mu\nu} dx^\mu dx^\nu =
a^2\left(1-\ptl_m P^m\right)^2d\eta^2-\gamma_{ij} (dx^i+ N^i
d\eta) (dx^j+ N^jd\eta),
\label{interv1}
\ee
where $x^\mu=\{\eta,\bm x\}$, $\eta$ is conformal time,
$\gamma_{ij}$ is a spatial metric, $a =\gamma^{1/6}$ is a locally
defined scale factor, and $\gamma=\det\gamma_{ij}$. The spatial
part of the interval (\ref{interv1}) reads as
\be
dl^2\equiv\gamma_{ij}dx^idx^j=a^2(\eta,\bm x)\tilde
\gamma_{ij}dx^idx^j,
\ee
where  $\tilde\gamma_{ij}=\gamma_{ij}/a^2$ is a matrix with the
unit determinant.

The interval (\ref {interv1}) is similar formally to the ADM one
\cite{adm}, but with the lapse function $N$ changed by the
expression $1-\ptl _ m P ^ m $, where $P ^ m $ is a
three-dimensional (relatively rotations) vector, and $\ptl _ m $
is a conventional particular derivative. Finaly, restrictions
$\ptl _ n(\ptl _ m N^m )=0$ and $\ptl _ n(\ptl _ m P^m )=0$ arise
because they are the Lagrange multipliers in FVT. Hamiltonian
$\mathcal H$ and momentum $\mathcal P_i$ constraints in the gauge
(\ref{interv1}) have the same form \cite{Cherkas2019} as in GR, so
in FVT, and obeys the constraint constraint algebra which gives
the constraint evolution equation \cite{Cherkas2019}. Let us write
it in the particular gauge $N=1$, $N^i=0$:
\bea
\ptl_\eta{\mathcal H}=\ptl_i\left(\tilde \gamma^{ij}\mathcal P_j\right),\label{5} \\
\ptl_\eta {\mathcal P_i}=\frac{1}{3}\ptl_i {\mathcal H}.\label{6}
\eea
 One could notice that the evolution of constraints governed by (\ref{5}), (\ref{6}) admit adding some constant to ${\mathcal H}$. Thus the constraint $\mathcal H$ not necessarily must be zero, but $\mathcal H=const$ is also admitted.  This fact explains why the Friedmann equation in FVT is satisfied up to some arbitrary constant. It is also interesting
how black holes look in the framework of FVT
\cite{cherkas2020eicheons}.

\section{Vacuum as a fluid}

\noindent Let us consider a quantum scalar field $\hat\phi (\eta,
\bm x) $ against a classical background of the uniform, flat,
expanding Universe with a space-time metric:
\be
ds^2\equiv g_{\mu\nu}dx^\mu dx^\nu
=a^2(\eta)\left(d\eta^2-\tilde\gamma_{ij}dx^idx^j\right),
\label{int}
\ee
where  $ \tilde\gamma _{ij}=\mbox{diag}\{1,1,1\}$ is a Euclidean
3-metric. At this moment, at least one fundamental scalar field is
known that is the Higgs boson \cite{Copeland2015}. Besides, as was
shown in \cite{Cherkas2007}, the gravitational waves contribute to
the vacuum energy density in the same manner as a scalar field.

The operators of the energy density and pressure of a scalar field
can be written as
\begin{eqnarray}
\hat\rho_\phi=\frac{1}{V}\int_V\left(\frac{\hat\phi^{\prime 2}}{2
a^2}+\frac{(\bm \nabla
\hat\phi)^2}{2 a^2}\right)d^3\bm r,\nonumber\\
\hat p_\phi=\frac{1}{V}\int_V\left(\frac{\hat\phi^{\prime 2}}{2
a^2}-\frac{(\bm \nabla \hat\phi)^2}{6 a^2}\right)d^3\bm r,
\end{eqnarray}
where $V$ is some normalizing volume which can be equaled unity.
Pressure and density of all kinds of matter define the scale
factor evolution of the flat Universe by the equations:
\begin{eqnarray}
-\frac{1}{2}M_p^2 a^{\prime 2}+\rho
a^4=const,\label{1}\\
M_p^2a^{\prime\prime}=(\rho-3 p) a^3,\label{2}
\end{eqnarray}
where the Planck mass $M_p=\sqrt{\frac{3}{4\pi G}}$. In the
frameworks of the FVT \cite{Cherkas2018}, the Friedmann equation
(\ref{1}) is satisfied up to some constant allowing to avoid the
problem of huge vacuum energy, which diverges as fourth degree of
momentum. A scalar field could be expanded over the plane wave
modes
 $\hat \phi(\bm
r)=\sum_{\bm k} \hat \phi_{\bm k} e^{i {\bm k}\bm r},$ which are
expressed through the creation and annihilation operators
\cite{Birrell1982}:
\begin{equation}
\hat \phi_{\bm k}=\hat {\mbox{a}}^+_{-\bm k}\chi_{k}^*(\eta)+\hat
{\mbox{a}}_{\bm k} \chi_{k}(\eta).
\end{equation}
The complex functions $\chi_k(\eta)$ satisfy the relations
\cite{Birrell1982}
\begin{eqnarray}
\chi^{\prime\prime}_k+k^2 \chi_k+2\frac{ a^\prime}{a}{
\chi^\prime}_k=0,\nonumber\\
a^2(\eta)(\chi_k \,{\chi_k^\prime}^*-\chi_k^*\,\chi_k^\prime)=i
\label{rel}
\end{eqnarray}
and can be found in the adiabatic approximation:
\begin{equation}
\chi_k(\eta)=\frac{\exp\left({-i \int_0^\eta
\sqrt{k^2-\frac{a''(\tau)}{a(\tau)}} \, d\tau}\right)}{\sqrt{2}
a(\eta) \sqrt[4]{k^2-\frac{a''(\eta)}{a(\eta)}}}.
\end{equation}
Let us calculate the mean vacuum energy density of a scalar field
\bea
\rho_v a^4=\frac{a^2}{2}\int \left(<0|{\hat\phi}^{\prime 2}|0>+
<0|(\bm \nabla{\hat\phi})^2|0>\right)d^3 \bm
r=\nonumber\\\frac{a^2}{2}\sum_{\bm k}<0|\hat \phi^\prime_{\bm
k}\hat\phi^\prime_{-{\bm k}}|0>+ k^2 <0|\hat\phi_{\bm
k}\hat\phi_{-{\bm k}}|0>=\frac{a^2}{2}\sum_{\bm k}
{\chi_k^\prime}^{*}\chi_k^\prime+k^2\chi_k^*
\chi_k\approx\nonumber\\\frac{1}{2}\frac{4\pi}{(2\pi)^3}\left(\frac{k_{max}^4}{4}
+\frac{k_{max}^2a^{\prime 2}}{4 a^2}+ O({a^\prime}^3)+O(a^\prime
a^{\prime\prime})+O(a^{\prime\prime\prime}+...)\right),~~~
\label{rh}
\eea
where it is implied that $a^{\prime 2}$, $a^{\prime\prime}$ has
the second-order of smallness, $a^{\prime 3}$,
$a^{\prime\prime}a^\prime$ are the third-order and so on
\cite{Cherkas2008}. Two first terms in Eq. (\ref {rh}) diverge as
4th and 2nd momentum degrees, respectively. The first term can be
omitted if the Friedmann equation (\ref{1}) is satisfied  up to
some constant. In the calculation of the second term, one could
use the ultra-violet cut-off $k_ {max}\sim M_p$ at the Planck mass
level \cite{Cherkas2007}.

One could ask why the $\bm k$-cutoff of comoving momentums is used
instead of, for instance, a cutoff of physical momentums related
to $\bm p=\bm k/a$ ($a$ is the universe scale factor)? The answer
could be that it is relatively simple to construct a theory with
the $\bm k$-cutoff, but it is challenging to introduce the $\bm
p$-cutoff fundamentally. For instance, merely considering gravity
on a lattice gives rather fundamental theory with comoving
momentums restricted by the period of a lattice.

Let us for definiteness to imply that the FVT model is considered
on some lattice in which the coordinate $\bm x$ takes discrete
values. As a result, one finds for the vacuum energy density:
\be
\rho_v=\frac{a^{\prime 2}}{2a^6}M_p^2S_0,
\label{r}
\ee
where
\[
S_0=\frac{1}{2 M_p^2}\sum_{\bm k}
\frac{1}{k}=\frac{1}{M_p^2(2\pi)^3}\int \frac{d^3\bm k}{2
k}=\frac{k_{max}^2}{8 \pi^2 M_p^2}.
\]
The vacuum energy density calculated is about of the critical
density of $M_p^2\mathcal H^2/2$. However, as a result of an
arbitrary constant on the right-hand side of Eq. (\ref{1}), the
concept of critical density loses its fundamental role as a
demarcating line between closed and opened Universes. Here, we
will consider a flat universe ad hoc.

\begin{figure}[th]
  \includegraphics[width=8.5cm]{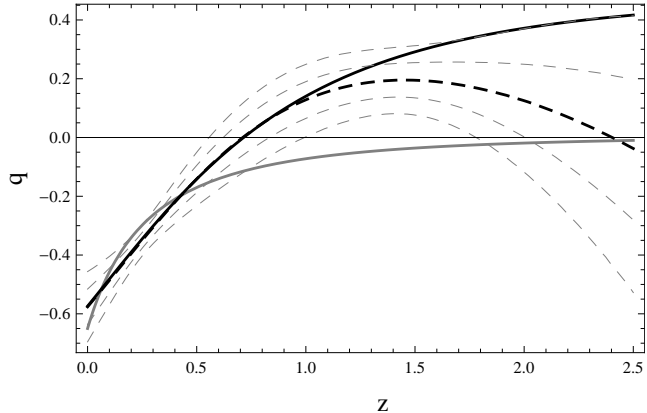}
\caption{\normalfont
  Deceleration parameter dependence on the redshift $z$. Solid
  black, gray solid and gray dashed curves correspond to the
  standard $\Lambda$CDM model, the vacuum domination model (\ref{ekk})
  of the present paper and the mean value of the observational data reconstruction \cite{Haridasu2018},
  respectively. Thin dashed curves point the 1$\sigma$ and  2$\sigma$ error channels of the reconstruction. }
\label{fig1}
\end{figure}

On the way to the vacuum ES finding, the next quantity has to be
calculated:
\begin{eqnarray}
<0|\hat \rho_\phi-3\hat p_\phi|0>=-\frac{1}{a^2}\int
\left(<0|{\hat\phi}^{\prime 2}|0>- <0|(\bm
\nabla{\hat\phi})^2|0>\right)d^3 \bm r=\nonumber\\
-\frac{1}{a^2}\sum_{\bm k}<0|\hat \phi^\prime_{\bm
k}\hat\phi^\prime_{-{\bm k}}|0>- k^2 <0|\hat\phi_{\bm
k}\hat\phi_{-{\bm k}}|0>=-\frac{1}{a^2}\sum_{\bm k}
a({\chi_k^\prime}^{*}\chi_k^\prime-k^2\chi_k^* \chi_k)
\approx\nonumber\\ \frac{1}{2a^6
}\left(a{a^{\prime\prime}}-{{a^\prime}^2}\right)\sum_{\bm
k}\frac{1}{k }+ O({a^\prime}^3)+O(a^\prime
a^{\prime\prime})+O(a^{\prime\prime\prime})+...,~~~~ \label{15}
\end{eqnarray}
which does not contain the terms $\sim k_ { max } ^4$. The
omission of items containing higher-order derivatives in
(\ref{15}) leads to
\bea
\rho_v-3p_v=\frac{1}{a^6
}\left(a{a^{\prime\prime}}-{{a^\prime}^2}\right)M_p^2S_0.
\label{3rp}
\eea
The vacuum pressure from Eqs. (\ref{3rp}) and (\ref{r}) is:
\be
p_v=\frac{M_p^2S_0}{a^6}\left(\frac{1}{2}a^{\prime
2}-\frac{1}{3}a^{\prime\prime}a\right).
\label{p}
\ee

\noindent It is easy to check that the vacuum energy density and
pressure determined by Eqs.  (\ref{r}), (\ref{p}) satisfy
\be
\rho_v^\prime+3\frac{a^\prime}{a}(\rho_v+p_v)=0.\label{3}
\ee
Eq. (\ref {3}) is one of the keystones describing the universe
evolution. It allows considering a vacuum as some ``fluid'' or
``substance'' with the well-defined dynamical ES, which can be
expressed explicitly.  For this goal, one needs to find a
dependence of the scale-factor $a$ on the conformal time. In a
more general case, when the universe filled with a cold dust-like
matter besides a vacuum, Eqs. (\ref{1}), (\ref {2}) take the form
\begin{eqnarray}
-\frac{1}{2}M_p^2 a^{\prime 2}+\rho_v
a^4+\frac{1}{2}M_p^2\Omega_m \mathcal H^2 a=const,\label{1a}\\
M_p^2a^{\prime\prime}=(\rho_v-3 p_v) a^3+\frac{1}{2}M_p^2\Omega_m
\mathcal H^2,\label{2a}
\end{eqnarray}
where $\Omega_m$ is a dimensionless constant characterizing the
density of matter and $\mathcal H$ is a value of a  of Hubble
constant at the present time $\eta=\eta_0$, when the universe
scale factor equals unity.

\begin{figure}[th]
\includegraphics[width=12.cm]{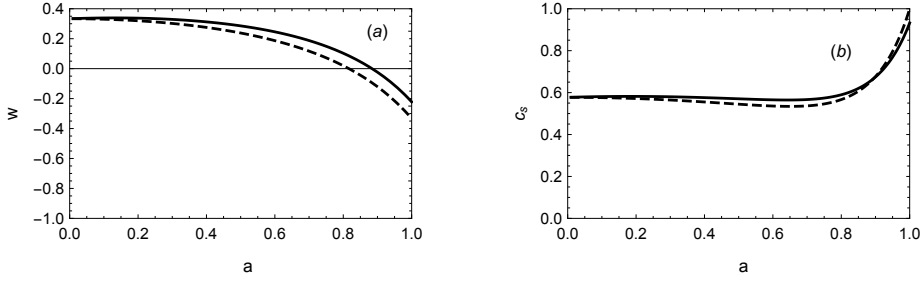}
\caption{\normalfont Dependencies of the vacuum ES (a) and the
velocity of the scalar ``sound waves'' (b) on the universe scale
factor. Solid line - $S_0=2.3$, $\Omega_m=0.3$, dashed line -
$S_0=2.3$, $\Omega_m=0.03$.}\label{fig2}
\end{figure}

The constant on the right hand side of Eq. (\ref{1a}) equals
$\frac {1}{ 2}\mathcal H^2M_p^2(S_0+\Omega_m-1)$, to satisfy
$a(\eta_0)=1$ and $a^\prime(\eta_0)=\mathcal H$. The resulting
Hubble constant dependence on scale factor is:
\begin{equation}
H(a)=\frac{\dot a}{a}=\frac{1}{a^2}\frac{da}{d\eta}=\frac{\mathcal
H}{a^2}\sqrt{\frac{S_0+\Omega_m-1-\Omega_m a}{S_0 a^{-2}-1}},
\label{ekk}
\end{equation}
where a dot denotes the differentiation on the cosmic time $dt=a
d\eta$. Finally, ES reduces to
\be
w_v=p_v/\rho_v={\left(1-\frac{2}{3}\frac{a^{\prime\prime}a}{a^{\prime
2}}\right)}=\frac{2 a^3 \Omega_m-3 a^2 (S_0+\Omega_m-1)+S
(S_0+\Omega_m-1)}{3 \left(a^2-S_0\right) ((a-1) \Omega_m-S_0+1)},
\label{vak}
\ee
where it is taken into account that $a^\prime=a^2 H(a)$ and
$a^{\prime\prime}=a^2 H(a)\frac{d}{da}\left(a^2 H(a)\right)$.

 Eq. (\ref{vak}) is  not singular up to the ``Big Rip'' (see, e.g., \cite{Ellis2012}) at
 $a=\sqrt{S_0}$, which will come in future if $S_0>1$.
 Explicit calculation of the vacuum energy density leads to
\be
 \rho_v=\frac{H^2 M_p^2 S_0 (S_0+\Omega_m-1-a \Omega_m)}{2 a^4
 \left(S_0-a^2\right)}.
 \ee

Let us remind that Eq. (\ref{3}) leads to $\rho a^{3(1+w)}=const$
for simple dependencies of $p=w\rho$, where $w=const$. In this
case, it is easy to write the Universe expansion law
$a^{\prime\prime}a^{-3w}=const$ from Eq. (\ref{2}). Thus, $a\sim
\eta^{2+3w}$, except for $w=-1/3$ when
\be
a\sim \exp\left(\mathcal H \eta\right).
\label{mmm}
\ee
The last case corresponds to the Milne's-like Universe
\cite{Milne1935}, i.e., to the exponential expansion in
``conformal time'' and to the linear one in ``cosmic time''
$dt=ad\eta$. It is necessary to remind that Milne's Universe is
spatially open, while we consider a flat Universe ad hoc.

For the vacuum ES $w\ne const$  the evolution becomes nontrivial.
Eqs. (\ref {r}), (\ref{p}) result in the defined ES, if the
expansion law is known, for example, $w_ {vac}\sim 1/3$ (i.e.,
radiation-like) for the Milne's-like Universe (\ref{mmm}).
However, if the vacuum has approximately the radiation-like ES at
some instant of time, it does not mean that the Universe expands
like the radiation dominated one. The point is that Eq. (\ref {2})
for the second derivative of the Universe scale factor contains
the expression $1-3w$. Thus, if $w$ is close to $1/3$ then
deviations from $w=1/3$ law play role and  determine the
evolution. For the pure radiation-dominant Universe, these
deviations are zero exactly, but for the vacuum dominated Universe
they turn out substantial that leads to the Milne's-like
expansion. Actually, $H(a)\sim 1/a$ at small scale factor
according to Eq. (\ref {ekk}), i.e., as it is for the Milne's
Universe.

The results of calculation of the deceleration parameter $q
(z)=-\frac{\ddot  a } {\dot a^2 } = \frac{1+z} {H}\, \frac { d
H(z) }{d z}-1 $ obtained from Eq. (\ref {ekk}) are shown in Fig.
\ref{fig1}. The Universe looks like the Milne's one for $z\geq2$,
where the deceleration parameter is close to zero, and then comes
to an acceleration phase. More general background model is
discussed in \cite{arx}.

Let us once more explain proximity to the Milne's law of the
Universe expansion at the simple particular case of $\Omega_m=0$
in which
\bea
\rho_v=\frac{H^2 M_p^2 S_0 (S_0-1)}{2 a^4
 \left(S_0-a^2\right)},\\
w_v=\frac{1}{3}-\frac{2 a^2}{3 (S_0- a^2)}.
 \eea
Eq. (\ref{2}) leads to
\be
\frac{M_p^2a^{\prime\prime}}{a}=\rho_v(1-3w_v)a^2=\frac{\mathcal
H^2M_p^2S_0(S_0-1)}{(S_0-a^2)^2},
\ee
and one has $\frac{a^{\prime\prime}}{a}\sim const$ at small $a$,
i.e.,  approximately the Milne's-like Universe. It is instructive
to compare that with the case of $w=-1/3$, $\rho=\frac{\mathcal
H^2 M_p^2 }{2 a^2}$ when $\frac{a^{\prime\prime}}{a}= const$,
i.e., exactly the Milne's expansion law.

Validity of Eq. (\ref{3}) allows describing a vacuum as some
absolutely elastic ``fluid'' with a ``sound-speed'':
\be
c_s^2=\frac{p^\prime}{\rho^\prime}=\frac{2 \left(5 a^5
\Omega_m-a^3 \Omega_m S_0+(7 a^2 S_0-9a^4-2S_0^2)
(\Omega_m+S_0-1)\right)}{3 \left(a^2-S_0\right) \left(5 a^3
\Omega_m-3 a \Omega_m S_0+(4 S_0-6 a^2)
   (\Omega_m+S_0-1)\right)}.
   \label{skor}
\ee

According to Eqs. (\ref {rh}), (\ref {15}), the waves of the
Planck-order frequency give the main contribution to the vacuum
pressure and density. These frequencies exceed the frequencies of
``vacuum sound waves''. That is, the local compressions/expansions
in a vacuum caused by these sound waves can be considered as the
expansion and collapse of some ``small universes''. Eq.
(\ref{skor}) implies  that the birth of particles from a vacuum,
which would increase its entropy, is negligible. This means that
an adiabatic vacuum is under consideration so that a ``fluid''
remains a vacuum during all the Universe evolution in the process
of the scalar sound waves propagation.

Fig. \ref{fig2} demonstrates that a dust-like pressureless matter
has a little impact on the vacuum ES and the corresponding sound
wave speed. The last increases from $1/\sqrt{3}$ up to some value
at the present time. With further expansion of the Universe, the
``sound speed'' exceeds the speed of light and tends to infinity
approximately at $a=\sqrt{\frac{2}{3} S_0 } $. That is, Eq.
(\ref{skor}) demonstrates that the ``Big Rip'' occurs earlier,
than it follows from ES (\ref{vak}). Regarding the speed of light
excess, it is difficult to say from the above empirical model
whether one deals with the physical effect \cite{Ellis2007} or
with a consequence of neglecting of a vacuum entropy.

Unlike the linearly expanding Universe with the ES of $w=-1/3$
\cite{John1996,Melia2015}, where the imaginary sound speeds are
possible, the sound speed is always positive in our case that
excludes the nonphysical solutions. Such nonphysical solutions can
be easily omitted in the analytical calculations
\cite{Cherkas2018}, but they remain the issue for the numerical
simulations.

\section{Masses and vacuum} A massless quantum field is considered
above. In the case of the massive fields, Pauli's idea could be
actual (see \cite{Visser18} and Refs.). That is a contribution of
masses to vacuum energy from bosons and fermions should compensate
each other. As a result, the main part of vacuum energy density is
\bea
\rho_v=\frac{1}{4
\pi^2a^4}\int_0^{k_{max}}k^2\sqrt{k^2+a^2m^2}dk\approx\frac{1}{16\pi^2}
\biggl(\frac{k_{max}^4}{a^4}\nonumber\\+\frac{ m^2
k_{max}^2}{a^2}+\frac{m^4}{8}
\left(1+2\ln\left(\frac{m^2a^2}{4k_{max}^2}\right)\right)\biggr).~~~~~~
\eea
 Simultaneously, three different principles could explain
why the main part of vacuum energy does not contribute to the
Universe evolution. The term $k_{max}^4$ is omitted in the FVT
gravity, where the energy reference level is arbitrary. The terms
$\sim m^4$ are a pure mass contribution to the vacuum density.
However, the condensates precipitate in the Standard Model of
Electroweak Interactions to generate masses itself. A density of
condensates has the same order of $m^4$. Overall compensation of
$m^4-$terms including condensates should be considered. This
problem stills unresolved yet, but it implies some unknown
symmetry mass generating potentials in Lagrangian allowing the
compensation with the accuracy at least of the order of $\sim
m_\nu^4$, where $m_\nu$ is the neutrino mass.
 The only
informative terms for particle physics are $\sim k_{max}^2m^2$,
which gives
\be
m_H^2+N_A m_{A}^2+6m_W^2+3m_Z^2=12m_t^2,
\ee
where the top quark mass is $m_t=173.2~GeV$, the Higgs boson mass
is $m_H=125~GeV$, the charged vector boson mass is $m_W=80,4~GeV$,
the neutral vector boson mass is $m_Z=91.2~GeV$, and $m_A$ is a
mass of unknown $A-$bosons contributing with the $N_A-$weight.
Thus, the physics of vacuum beyond the Milne-like stage of the
Universe expansion anticipates unknown bosons: a single boson
$m_A\sim 530~Gev$, or, for instance, four bosons $m_A\sim
265~Gev$.

\section{Formation of matter structures in the universe}

The ES and the scalar waves speed in a vacuum found in the
previous section could serve as the basis for the description of
the perturbations evolution of vacuum, radiation, and matter in
the expanding Universe. As was mentioned above, the Friedmann
equation is satisfied up to some constant in the FVT that allows
choosing an arbitrary reference level of the vacuum energy
density. Briefly, the FVT theory is based on the standard
Einstein-Hilbert action which is varied not over all the possible
metrics, but over some restricted class of them
\cite{Cherkas2018}. As a result, the Hamiltonian constraint turns
out to be weaker than that in GR. Perturbations of the metric of
the expanding Universe looks as
\begin{figure}[th]
  \includegraphics[width=12.5cm]{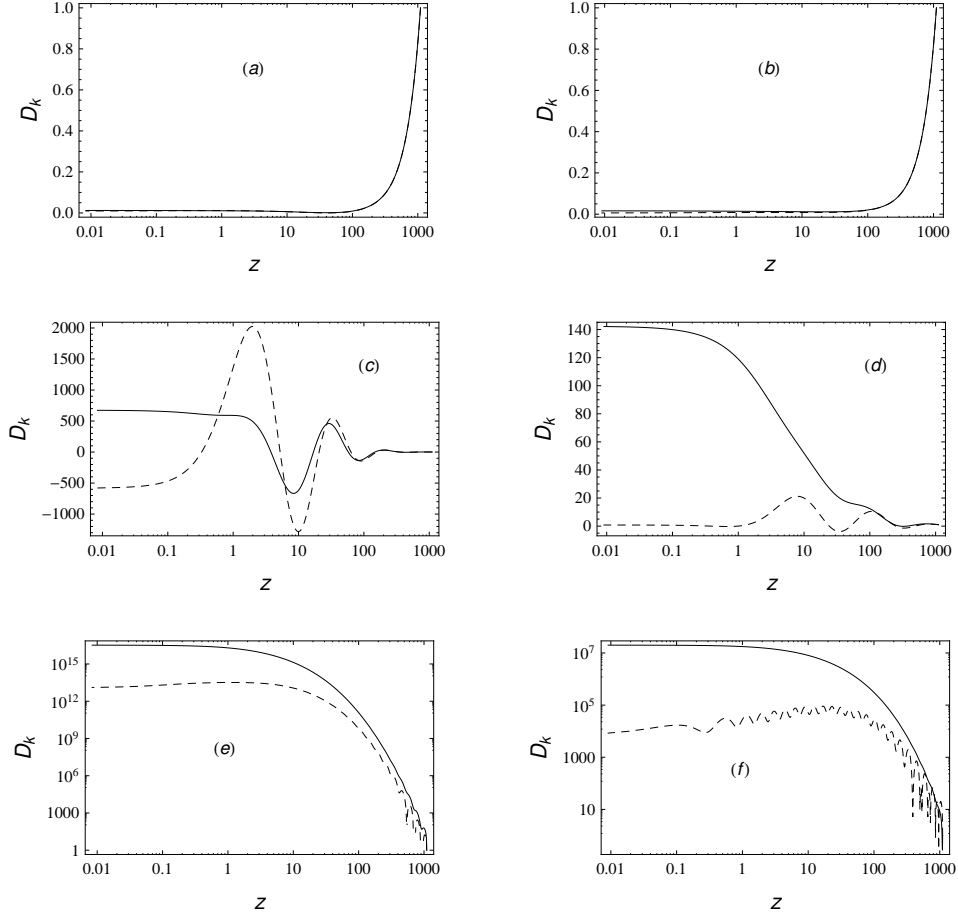}\\
  \caption{\normalfont
 The inhomogeneity growth factors for the different perturbation wave numbers: a,b- $k/h=0.0001 ~\mbox{Mpc}^{-1}$, c,d-
  $k/h=0.001~ \mbox{Mpc}^{-1}$, e,f- $k/h=0.01 ~\mbox{Mpc}^{-1}$. Solid curves correspond to a matter, dashed
  curves correspond to a vacuum. $\Omega_m=0.3$ for (a,c,e) and $\Omega_m=0.03$ for (b,d,f), respectively.}
  \label{fig3}
\end{figure}
\be
ds^2=a(\eta)^2(1+2A)\left(d\eta^2-\left(\left(1+\frac{1}{3}\sum_{m=1}^3
\ptl_m^2F\right)\delta_{ij}-\ptl_i\ptl_jF\right)dx^idx^j\right).
\label{int1}
\ee
The metric (\ref{int1}) belongs to the class of metrics
(\ref{interv1}) admissible in FVT \cite{Cherkas2018}.

Perturbations of density, pressure and 4-velocity of every
$c$-fluid are considered as $\rho_c(\eta,\bm x)
 =\rho_c(\eta)+\delta \rho_c(\eta,\bm
x)$, $p_c(\eta,\bm x)=p_c(\eta)+\delta p_c(\eta,\bm x)$,
\be
u^{\mu}_c=\frac{1}{a(\eta)}\{(1-A),\bm \nabla v_c(\eta,\bm x)\},
\label{vr}
\ee
where $v_c$ is a velocity potential.  The resulting system of
equations was obtained for the Fourier components of $\delta
\rho_c(\eta,\bm x)=\sum_{\bm k}\delta\rho_{c\bm k}(\eta)e^{i \bm k
\bm x }...$
\bea
-6 A_{\bm k}'+6 A_{\bm k} \alpha '+k^2 F_{\bm k}'+\frac{18}{M_p^2}
e^{2 \alpha } \sum_cV_{c \bm k}
=0,\label{con1}\\
 -18 \alpha ' A_{\bm k}'-18 A_{\bm k} \alpha ^{\prime 2}-6 k^2A_{\bm k}
+k^4F_{\bm k} +\frac{18}{M_p^2} e^{2 \alpha } \sum_c \delta
\rho_{c\bm k}+4 A_{\bm k}\, \rho_c
=0,\label{con2}\\
-12 A_{\bm k}-3 \left(F_{\bm k}''+2 \alpha ' F_{\bm k}'\right)+k^2
F_{\bm k}=0,\\
-9 \left(A_{\bm k}''+2 \alpha' A_{\bm k}'\right)-18 A_{\bm k}
\alpha''-18 A_{\bm k} \alpha^{\prime 2}-9 k^2 A_{\bm k} +k^4
F_{\bm k} \nonumber\\-\frac{9}{M_p^2} e^{2 \alpha }\sum_c 4 A_{\bm
k} (3 p_c-\rho_c)+3 \delta p_{c\bm k}-\delta \rho_{c\bm k}=0,\\-3
\alpha' ( \delta p_{c\bm k}+\delta \rho_{c\bm k})-3 A_{\bm k}'
(\rho_{c}+p_{c})-\delta \rho_{c\bm k}'+k^2 V_{c\bm k}=0,\\
(\rho_{c}+p_{c}) A_{\bm k}+4 V_{c\bm k} \alpha '+\delta p_{c\bm
k}+V_{c\bm k}'=0,\label{lasteq}
\eea
where  $V_{c}=(p_c+\rho_c)v_{c}$ corresponds to every kind of a
fluid.

Let us remind that ``gauge invariant'' potentials are usually
under consideration in GR that corresponds to the metric
\be
ds^2=a^2(\eta)\left((1+2\Phi(\eta,\bm
x))d\eta^2-\left(1-2\Psi(\eta,\bm
x)\right)\delta_{ij}dx^idx^j\right),
\label{muhmet}
\ee
as well as the ``gauge invariant'' density contrasts and the
velocity potentials:
\bea
\tilde \delta_{c\bm k}(\eta )= \frac{\delta \rho_{c\bm k}(\eta
)}{\rho_c(\eta )}+ \frac{\rho_c^\prime(\eta )}{2\rho_c(\eta )}
F_{\bm k}^\prime(\eta ),~~~~~ \tilde v_{c\bm k}=\frac{V_{c\bm
k}(\eta )}{\rho_c(\eta )+p_c(\eta
)}-\frac{F^\prime_{\bm k}(\eta )}{2},\nonumber\\
\Phi_{\bm k}(\eta)=A_{\bm k}(\eta)+\frac{a'(\eta ) F_{\bm k}'(\eta
)+a(\eta ) F_{\bm k}''(\eta )}{2 a(\eta )}, \nonumber\\
\Psi_{\bm k}(\eta)=-\frac{a'(\eta ) F_{\bm k}'(\eta )}{2 a(\eta
)}-A_{\bm k}(\eta )+\frac{1}{6} k^2 F_{\bm k}(\eta).
\label{invq}
\eea

If the Friedmann equation is satisfied exactly, Eqs. (\ref{con1})
- (\ref {lasteq}) can be rewritten in the terms of ``invariant''
quantities that results in the known equations \cite{Dodelson2003,
Mukhanov2005}.  However, if Friedman equation is satisfied up to
only some constant, the fundamental system is (\ref{con1}) - (\ref
{lasteq}).  In this case, it is impossible to rewrite this system
in the terms of invariant variables because of the metric (\ref
{muhmet}) does not belong to a class of metrics regarding which
the action varies in the FVT gravity \cite{Cherkas2017}.

Here, the authors consider a linear evolution of the
inhomogeneities of pressureless matter and vacuum beyond ``the
last scattering surface'', when radiation decouples with the
matter, and the Universe structure starts to develop
\cite{Lukash2010,Dolgov2019}. As is known, the anisotropy of CMB
imprints the degree of spatial inhomogeneity of a baryon-photon
plasma at the last scattering surface.

After decoupling, the inhomogeneities growth with the Universe
evolution results in the formation of structures such as galaxies,
clusters, and superclusters ($D_k\geq10^{10}, 10^7$, and $10^4$,
respectively, \cite{Longair2008}. Let's calculate the
inhomogeneity growth factor (the ``density contrast'' factor):
\be
D_k(z)=\tilde\delta_k(z)/\tilde\delta_k(1100),
\label{D}
\ee
where $z=1100$ is the redshift corresponding approximately to the
last scattering surface. Eq. (\ref{D}) contains the ``invariant''
variables, i.e., the calculation is performed in the reference
system (\ref{int1}), but one turns finally to the expressions
(\ref{invq}) which are the reference-frame invariant.

As is seen from Fig. \ref{fig3}, \emph{a, b}, the inhomogeneities
at the extra-large scale decrease for both matter and vacuum. At
the intermediate scale (Fig. \ref{fig3}, \emph{c, d}), vacuum
decouples with matter in a sense that its perturbations grows
slower. The value of the growth factor suggests that the  linear
theory is still valid, because the typical value of
inhomogeneities at the last scattering surface is estimated as
$10^{-4}-10^{-5}$. Multiplying these value by the grows factor
results in quantity less than unity. At smaller scales of the
order of galaxy clusters shown in Fig. \ref{fig3}, \emph{e, f},
the inhomogeneities enter into a nonlinear regime. In the standard
$\Lambda$CDM model this scale is ``slightly''-nonlinear
($D_k\leq10^4 - 10^5$), but it is strongly nonlinear in our model.
We conjecture that as an evidence of early and more intensive
structure formation demonstrated by the modern observational data
\cite{Melia2014,Oesch2016,Waters2016,Dolgov2019}.

At smaller scales one might conjecture that such vacuum
clusterization would be considered as a ``dark-matter halo''
formation, but such nonlinear regimes are far beyond the scope of
the present paper considering only linear perturbations evolution.
The above calculations are performed for two values of the
pressureless matter $\Omega_m=0.03$, as in the standard model, and
$\Omega_m=0.3$. The last value is preferable for Milne's-like
Universe because the nucleosynthesis in linearly coasting
cosmology demands this value of baryonic matter to provide
necessary amount of helium \cite{Lewis2016,Singh2018}.

\section{Conclusion}

It is shown that the description of a vacuum as some elastic
medium (``fluid'') leads to the ES with the defined speed of
scalar ``sound waves''. Such a representation can be considered as
a basis for the precision cosmology of the Milne's-type Universe
\cite{John1996,Melia2015,Dev2002}, with expansion close to linear.
Although the horizon problem is absent for such a model, the
Hubble constant $\mathcal H$ plays a role of a typical scale for
the evolution of perturbations. In particular, the perturbations
with a wave number $k<\mathcal H$ decrease during the Universe
evolution, while the perturbations with $k> \mathcal H$ increase.
According to numerical estimations, there is no need in the dark
matter for perturbations growth, because of the perturbations
increase intensively at the small scales and enter into the
nonlinear regime. It seems that the Milne's viewpoint on the
necessity to proceed from a ``cosmological picture'' and
``descent'' to a local theory of gravitation still could be more
relevant than it usually considered. Namely one should describe
the physical and cosmological properties of vacuum fluctuations
first, and only then introduce lacking pieces like dark matter and
energy.

Despite the active latest debates on the Milne's-like cosmologies
(``freely coasting universe'', ``$R_h=c t$-universe,'' etc.), the
discourse is staying on the natural philosophy level until now.
This paper aims to divert this discussion into physical context.
Namely, the vacuum ES unifying the dark energy/matter and the
system of equations for the perturbations evolution provides the
necessary calculational paradigm for the quantitative comparison
with the standard model. One has to note that the nonlinear
evolution of perturbations is much more tricky for analysis
because it could require the consideration of nonlinear operators
evolution for the energy density of quantized fields.

\bibliographystyle{h-physrev5.bst}

\bibliography{structure}

\end{document}